\documentclass[showpacs,prb,superscriptaddress,twocolumn]{revtex4}
\usepackage{graphicx}
\usepackage{amsmath}
\usepackage{amssymb}

\begin{document}

\title{Near-Zero Moment Ferromagnetism in the Semiconductor SmN}
\author{C. Meyer}
  \affiliation{The MacDiarmid Institute for Advanced Materials and Nanotechnology,
    School of Chemical and Physical Sciences,
    Victoria University,
    PO Box 600, Wellington 6140, New Zealand}
  \affiliation{Institut N\'eel, CNRS-UJF, BP 166, 38042 Grenoble, France}
\author{B. J. Ruck} \email{ben.ruck@vuw.ac.nz}
  \affiliation{The MacDiarmid Institute for Advanced Materials and Nanotechnology,
    School of Chemical and Physical Sciences,
    Victoria University,
    PO Box 600, Wellington 6140, New Zealand}
\author{J. Zhong}
  \affiliation{The MacDiarmid Institute for Advanced Materials and Nanotechnology,
    School of Chemical and Physical Sciences,
    Victoria University,
    PO Box 600, Wellington 6140, New Zealand}
\author{S. Granville}
  \affiliation{The MacDiarmid Institute for Advanced Materials and Nanotechnology,
    School of Chemical and Physical Sciences,
    Victoria University,
    PO Box 600, Wellington 6140, New Zealand}
\author{A. R. H. Preston}
  \affiliation{The MacDiarmid Institute for Advanced Materials and Nanotechnology,
    School of Chemical and Physical Sciences,
    Victoria University,
    PO Box 600, Wellington 6140, New Zealand}
\author{G. V. M. Williams}
  \affiliation{The MacDiarmid Institute for Advanced Materials and Nanotechnology,
    Industrial Research Ltd.,
    PO Box 31310, Lower Hutt 5040, New Zealand}
\author{H. J. Trodahl}
  \affiliation{The MacDiarmid Institute for Advanced Materials and Nanotechnology,
    School of Chemical and Physical Sciences,
    Victoria University,
    PO Box 600, Wellington 6140, New Zealand}
  \affiliation{Ceramics Laboratory, EPFL-Swiss Federal Institute of
  Technology, Lausanne 1015, Switzerland}

\begin{abstract}
The magnetic behaviour of SmN has been investigated in
stoichiometric polycrystalline films. All samples show ferromagnetic
order with Curie temperature ($T_C$) of $27 \pm 3$~K, evidenced by
the occurrence of hysteresis below $T_C$. The ferromagnetic state is
characterised by a very small moment and a large coercive field,
exceeding even the maximum applied field of $6$~T below about
$15$~K. The residual magnetisation at $2$~K, measured after cooling
in the maximum field, is $0.035~\mu_B$ per Sm. Such a remarkably
small moment results from a near cancellation of the spin and
orbital contributions for Sm$^{+3}$ in SmN. Coupling to an applied
field is therefore weak, explaining the huge coercive field . The
susceptibility in the paramagnetic phase shows
temperature-independent Van~Vleck and Curie-Weiss contributions. The
Van~Vleck contribution is in quantitative agreement with the
field-induced admixture of the $J=\frac{7}{2}$ excited state and the
$\frac{5}{2}$ ground state. The Curie-Weiss contribution returns a
Curie temperature that agrees with the onset of ferromagnetic
hysteresis, and a conventional paramagnetic moment with an effective
moment of $0.4~\mu_B$ per Sm ion, in agreement with expectations for
the crystal-field modified effective moment on the Sm$^{+3}$ ions.
\end{abstract}

\pacs{75.50.Pp, 75.30.Cr, 74.70.Ak}

\date{\today}

\maketitle

\section{Introduction}

The rare-earth nitrides (RN, R = rare-earth atom) have gained
attention recently as simple (NaCl) structures for which the
influence of strong correlations on the electronic band structures
can be treated with some confidence \cite{larson2007esr}. In
parallel with theoretical advances there has developed an
experimental interest in the growth and passivation of thin films
\cite{leuenberger2005gtf,granville2006sgs}. This interplay of theory
and experiment has revealed a number of interesting properties of
both fundamental and technological importance. Firstly the
ambient-temperature paramagnetic phase has a narrow indirect gap
that varies systematically across the series. Secondly in the
ferromagnetic state (which may be the ground state for them all)
they are predicted to have spin-polarised carriers, opening the
potential to doped spintronic structures. Early data on these
compounds were plagued by a lack of stoichiometric reproducibility
and a rapid degradation of the RN under atmosphere
\cite{hullinger1978hot,vogt1993hot}. The magnetic properties in
particular are very sensitive to nitrogen vacancies and oxygen
impurities, which are difficult to control in these materials. Even
the exchange interactions between the rare earth $4f$ spins are not
well understood, though a number of theoretical models have been
proposed \cite{duan2007}. The exchange is usually described to
operate within two competitive channels of superexchange via the
nitrogen atom. The nearest-neighbor (nn) interaction is configured
at $90^\circ$ and is believed to be ferromagnetic. It strongly
depends on the carrier concentration and becomes dominant when RKKY
indirect interactions take place via the polarisation of conduction
electrons. The next-nearest-neighbor (nnn) interaction is configured
at $180^\circ$, is antiferromagnetic and in principle dominant for
non-metals. Note that the existence of a ferromagnetic order in
semiconducting rare-earth nitrides implies that a ferromagnetic
interaction dominates even in the absence of free carriers
\cite{duan2007}.

The most thoroughly studied of these compounds is GdN, which has a
half-filled $4f$ shell with the maximum $\frac{7}{2}$ net spin and
zero net orbital angular momentum. It is ferromagnetic below $70$~K,
with a saturation moment of $7 \mu_B$ and is an indirect-gap
semiconductor with an optical gap of $1.3$~eV in the paramagnetic
phase, reduced to $0.9$~eV below the Curie temperature
\cite{trodahl2007fro}. For the lighter rare-earths, Hund's rules
specify that $L$ is anti-parallel to $S$ in the ground state. Within
that scenario SmN is of special interest; Sm$^{+3}$, with two
electrons below half-filling, has $S = \frac{5}{2}$ and $L = 5$ and
a net magnetic moment given by $M \approx (L_z+2S_z)\mu_B \ll
\mu_B$. There is thus the potential for Sm compounds to condense
into a ferromagnetic phase in which the spins are ferromagnetically
ordered, but with their spin moment nearly cancelled by an opposing
orbital moment. Moments substantially smaller than the free-ion
moment of $0.71 \mu_B$ are not unknown in ferromagnetic Sm compounds
\cite{givord1979uff,stewart1974fos,wijn1973eoc,adachi1999sot,adachi2001zmf,ahn2007pra},
but its occurrence in a semiconductor has to our knowledge not been
reported previously. Such a material offers special advantages for
spintronics: (i) it can inject spin-polarised electrons into a
conventional semiconductor without the deleterious effects of a
fringe magnetic field \cite{nie2008phm}, and (ii) in principle it
can form field-free, fully spin-polarised electronically active
structures.

Early magnetic measurements suggested that SmN was antiferromagnetic
below $20$~K \cite{hullinger1978hot,vogt1993hot}, but this was not
confirmed by neutron diffraction \cite{moon1979mpo}, suggesting that
it might indeed be ferromagnetic but with near-cancellation between
the spin and orbital moments. More recently we reported clear
ferromagnetism in GdN \cite{granville2006sgs, trodahl2007fro} and
DyN \cite{preston2007bso}, but somewhat weaker ferromagnetic
evidence in SmN \cite{preston2007bso}. The resistivities of the
films show them all to be semiconductors, with the expected anomaly
at $T_C$ signalling a narrowed gap in the ferromagnetic state. In
the present work we report magnetic experiments performed on thicker
SmN films, seeking to resolve the uncertainties concerning the
magnetic state of this unusual material, and clear picture of
near-zero moment ferromagnetism emerges.

\section{Experimental details}

SmN films were grown in a vacuum chamber pumped to a base pressure
in the $10^{-9}$~mbar range. Sm metal was evaporated in the presence
of an atmosphere of pure N$_2$ gas at a pressure of $10^{-4}$~mbar;
the growth conditions, the structure and stoichiometry have been
reported previously \cite{granville2006sgs, trodahl2007fro}. For the
present measurements the films were deposited on Si substrates
covered by their natural oxide. X-ray diffraction exhibits the Bragg
peaks of only the rock salt cubic structure and establishes the
films as untextured polycrystalline with an average crystal grain
size of $10$~nm. The lattice parameter ($5.07$~{\AA}) is consistent
with the previous data for the rare earth nitride series, confirming
that samarium is trivalent. All ex-situ measurements are performed
on films protected by a cap layer of nanocrystalline GaN.
Conductivity and x-ray spectroscopies on the films have established
them to be semiconductors in both the ambient-temperature state and
to $4$~K in the magnetically ordered low temperature
state~\cite{preston2007bso}.

Three films of differing thickness ($300-400$~nm) were used for
magnetic measurements reported here. These magnetic properties were
investigated with a SQUID magnetometer (Quantum Design MPMS) working
up to a maximum applied field of $6$~Tesla. All experiments were
performed with the magnetic field applied parallel to the film
plane. The films were prepared in parallel on both thick
($400~\mu$m) and thin ($100~\mu$m) Si substrates in order to apply
more reliable corrections for substrate signals, which are of the
same order of magnitude as the SmN signal. Si is diamagnetic and the
susceptibility is supposed to be temperature independent. The
susceptibility measured on the uncovered Si substrates is in
agreement with the theoretical susceptibility $\chi(Si)=-3.4\times
10^{-6}$ at room temperature within a $5\%$ error. This Si signal is
characterized well enough to permit a trouble-free correction in the
data shown below.

The magnetic signal from the capping layer is somewhat more
problematic. GaN is weakly paramagnetic mainly because self-doping
is provided by N vacancies, for which we collected reference data
from a measurement on a Si substrate covered with a GaN layer of the
same thickness. In a phenomenological approach, we have fit the
temperature variation of the magnetisation using a theoretical
calculation by Sonder and Schweinler \cite{sonder1960} predicting a
modified Curie law for the susceptibility of interacting donor
centres in doped semi-conductors: $\chi=\frac{C}{T^{(1-\alpha)}}$.
In this model the parameter $\alpha$ is proportional to the donor
concentration $n_d$. We find $C=4.05\times 10^{-3}$ and
$\alpha=0.86$, but uncertainty about the donor concentration
prevents further analysis. We note that this approach has been used
to explain the susceptibility of a zinc-blende GaN thin film
\cite{fanciulli95}. The authors derive $\alpha\simeq0.8$ for which
they estimate $n_d=2\times 10^{17}$ cm$^{-3}$, in agreement with the
experimental value attributed to nitrogen vacancies.

However, the susceptibility of GaN is very sensitive to N vacancy
concentration \cite{fanciulli95}. There is some variation in the
susceptibility of the GaN capping layers associated with minor
differences in stoichiometry resulting from the ion-assisted
deposition process. Nevertheless we are able to subtract the capping
layer signal to reasonable accuracy, so the evidence of the
remarkable magnetic behaviour of SmN is not strongly affected by
this uncertainty. Below we present the data both uncorrected and
after correction for the substrate/capping layer signals.

\section{Results}

\begin{figure}
\centering{
  \includegraphics[width=8cm]{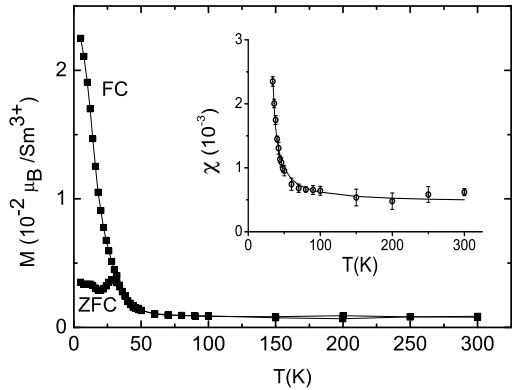}
} \caption{ Temperature dependence of the magnetisation of SmN after
cooling without (ZFC) and with an applied field (FC) of $0.5$~T
(sample \textit{I}). The inset shows the result of the fit of the FC
susceptibility ($\chi = M/H$) in a Van~Vleck approach.} \label{fig1}
\end{figure}

Figure \ref{fig1} shows the temperature variation of the
magnetisation of sample I in an applied field of $0.5$~T, after
cooling in zero field (ZFC) or in $0.5$~T (FC). The curves are
superimposed down to $30$~K, below which a sharp increase of the FC
curve denotes ferromagnetic order with a spontaneous magnetic
moment. The same behaviour has been confirmed in an applied field as
low as $5$~mT. In the ZFC curve the increase of the magnetisation
below $20$~K is assigned to the GaN cap layer. The transition
temperature estimated from the maximum of the ZFC curve is found at
$T_C = 30~\pm~2$~K. The other two films gave Curie temperatures of
$24~\pm~2$~K and $26~\pm~2$~K. We have not been able to relate these
small differences to the films' compositions or structures, so we
quote $T_C = 27 \pm 3$~K. Note in this regard that it is known that
N vacancies lower the Curie temperature in GdN \cite{cutler1975sam}.
Our films are close to stoichiometric, but absolute measurements of
the composition have an accuracy of about $5\%$. It is notable that
earlier heat capacity \cite{stutius1969tsh} and magnetisation
\cite{busch1965ioc} measurements suggested Curie temperatures in the
$15-20$~K range; they were likely performed on N deficient samples.

\subsection{Ferromagnetic state}

\begin{figure}
\centering{
  \includegraphics[width=10cm]{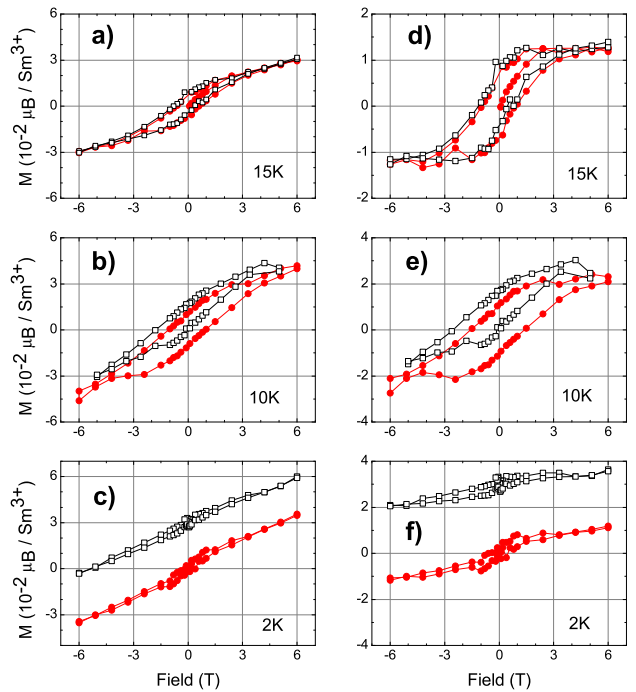}
} \caption{ (color online) Magnetisation loops of SmN\textit{II}
after cooling the film in zero-field (red full circles) and after
cooling under $6$~T (open black squares). Left: correction for the
Si substrate only at (a)~$15$~K, (b)~$10$~K and (c)~$2$~K. Right:
additional correction for the paramagnetic signal of the GaN cap
layer according to the curve shown on Figure \ref{fig3} at
(d)~$15$~K, (e)~$10$~K and (f)~$2$~K .} \label{fig2}
\end{figure}

Above $T_C$ the magnetisation $M$ is linear in field $B$ with a
paramagnetic susceptibility that includes both Curie-Weiss and
temperature-independent Van~Vleck contributions as will be discussed
in the following section. At $T_C$ $M(B)$ becomes nonlinear, and an
hysteresis loop develops. The evolution of the loop as the
temperature is lowered is shown in Figure \ref{fig2}. Here the left
side [Fig.~\ref{fig2}(a, b, c)] represents the loops uncorrected for
GaN while in the right side [Fig.~\ref{fig2}(d, e, f)] the
correction is included. We first focus on the loops measured after
cooling the films in zero field (red full dots). At $15$~K
[Fig.~\ref{fig2}(a)] the saturation is achieved at $3$ T. Beyond the
irreversibility point, the magnetisation is linear in field, due to
the paramagnetic contribution of GaN. After correction for GaN the
loop exhibits a coercive field of $0.9$~T [Fig.~\ref{fig2}(d)]. At
$10$~K [Fig.~\ref{fig2}(b)] the reversible part at high field is
missing; the closing field lies above the $6$~T maximum available.
This behaviour is exactly as expected when the maximum applied field
is insufficient to achieve the reversal of the moments, so that only
minor loops are measured, signaling a magnetocrystalline anisotropy
that grows at lower temperature \cite{callen1960am}. At $5$~K the
loop shows only a very small opening and at $2$~K
[Fig.~\ref{fig2}(c)] the hysteresis has completely disappeared, so
that only the GaN paramagnetic contribution is seen in the ZFC data.
Clearly the magnetic field necessary to even initiate the reversal
of ferromagnetic domains is higher than $6$~T at these temperatures,
and the magnetisation process is dominated by the reversible
paramagnetic contribution. Hysteretic behaviour can nonetheless be
confirmed at these temperatures by cooling in the presence of the
maximum field of $6$~T to prepare the film in a magnetised state.
Thus Figure \ref{fig2} compares the hysteresis patterns obtained at
$15$~K, $10$~K and $2$~K after zero-field cooling (red full dots)
and after cooling in $6$~T (black open squares). The loops are
superimposed at $15$~K [Fig.~\ref{fig2}(a,d)], but at the lower
temperatures the patterns are shifted from one another
[Fig.~\ref{fig2}(b,e) and Fig.~\ref{fig2}(c,f)]. Exactly the same
behaviour is observed when cooling the system under $-6$~T, though
with the shifts found in the opposite sense. The results confirm
that the coercive and saturation fields are larger than $6$~T at
these temperatures.

In addition a very small spontaneous moment is observed on the
saturated loops; the temperature dependence of the magnetic moment
down to $0$~K is interesting to evaluate. At $15$~K, $M_S = 0.012
\mu_B/$Sm, compared to $0.008 \mu_B/$Sm at remanence when the field
is reduced to zero. The difference is due in part to single-domain
crystallites relaxing to an easy-axis magnetisation in the remanent
state, leaving the moments distributed in a cone about the field
direction. The easy axis in SmN is as yet unknown, but if it lies
along one of the high-symmetry directions $\langle100\rangle$,
$\langle110\rangle$ or $\langle111\rangle$, the field-parallel
component of the moment is reduced by about $15\%$, explaining about
one half of the measured reduction from the saturation to the
remanence. We conclude that the remnant moment provides a reasonable
lower limit for the single-domain spontaneous magnetisation. To
refine the estimate we have performed experiments in which the
material is prepared in the magnetised state by cooling in $6$~T,
followed by measurements of the magnetisation in a small field of
$20$~mT. The resulting remnant magnetisation drops to zero at the
Curie temperature from a zero-temperature magnetisation of $0.030
\pm 0.006 \mu_B/$Sm averaged over the three films. Assuming that
this value represents somewhat less than $85\%$ of the spontaneous
moment we quote that moment as $0.035 \pm 0.010 \mu_B/$Sm at the
lowest temperature. Such a small moment explains the null result in
the early neutron search for ferromagnetic order \cite{moon1979mpo},
and is in agreement with the near cancellation of spin and orbital
moments suggested by Larson~\textit{et~al.}~\cite{larson2007esr}.
The saturation field is large, rising above our $6$~T maximum
available field below $15$~K. It is important in this regard to note
that the ferromagnetic state, with its very small moment, couples
relatively weakly with the magnetic field.

\subsection{Paramagnetic state}

Remarkably, the small moment found in the ferromagnetic phase is not
carried across the transition, rather the paramagnetic behaviour of
SmN can be fully understood within the established description of Sm
in the crystalline environment. We start by recalling that the
ground state configuration of the Sm$^{3+}$ free ion is
$^{6}H_{\frac{5}{2}}$ with $L=5$, $S=\frac{5}{2}$, and
$J=\frac{5}{2}$. The Land\'e factor is $g_J = \frac{2}{7}$ and the
magnetic moment is $\mu = 0.71 \mu_B$. The first excited multiplet
$J=\frac{7}{2}$ is not thermally populated, but a paramagnetic
moment will partly arise from an admixture of the multiplets induced
by the applied magnetic field. As usually observed in trivalent Sm
compounds, the reciprocal susceptibility is therefore not linear in
temperature, preventing a Curie-Weiss-like analysis.

With the inclusion of the $J = \frac{7}{2}$ admixture in the
$\frac{5}{2}$ ground state, the paramagnetic susceptibility of
Sm$^{3+}$ compounds is reasonably well described by the Van~Vleck
approach, with \cite{stewart1972pso}

\begin{equation}
    \chi = \chi_0 + \frac{C}{T-\Theta_p}.
  \label{eq:1}
\end{equation}

The second term is the conventional Curie-Weiss susceptibility
involving the effective magnetic moment $\mu_{\mathrm{eff}}$ for
$J=\frac{5}{2}$, with $C = \mu_0 N \mu_{\mathrm{eff}}^2 / 3 k_B$,
where $N$ is the Sm ion density. The Van~Vleck term $\chi_0$ depends
on the energy difference $\Delta E \simeq 1500$~K between the two
lowest $J (\frac{5}{2}, \frac{7}{2})$ multiplets \cite{wijn1973eoc,
stewart1972pso}:

\begin{equation}
    \chi_0 = \frac{\mu_0 N {\mu_B}^2}{k_B} \cdot \frac{20}{7 \Delta E}.
    \label{eq:2}
\end{equation}

\begin{figure}
\centering{
  \includegraphics[width=8cm]{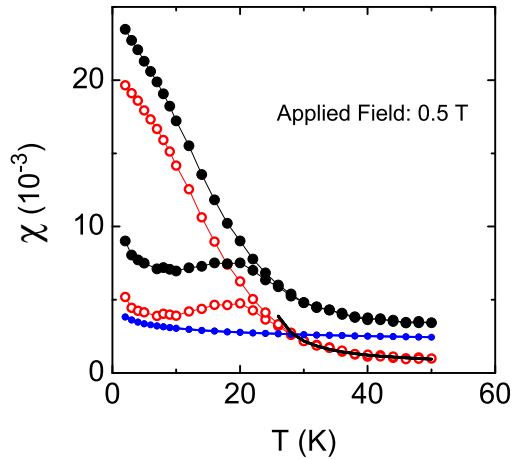}
} \caption{ (color online) Susceptibility ($\chi = M/H$) versus
temperature (sample\textit{II}), before (full black circles) and
after (open red circles)the correction for the GaN cap layer. The
GaN signal used for the correction is shown in small blue dots. The
full black line is the Van Vleck fit of the susceptibility to the
fully corrected data, setting the multiplet separation $\Delta
E=1500$~K.} \label{fig3}
\end{figure}

We have used Equation~(\ref{eq:1}) to fit the measured paramagnetic
susceptibility curves above $T_C$ as shown in the inset of
Fig.~\ref{eq:1} for the magnetisation curve obtained under $0.5$~T.
It can be seen that the free-ion Van~Vleck term, with no adjustable
parameters, provides an excellent fit to the temperature-independent
tail at high temperatures. The diverging contribution below $100$~K
yields a Curie temperature $\theta_P = 28 \pm 1$~K, in excellent
agreement with the value of $30 \pm 2$~K where this film showed a
cusp in the ZFC magnetisation curve. Similar agreement has been
found also with the other two films, as can be seen on Figure
\ref{fig3} for sample \textit{II} . A sense of the relative strength
of the contributions to the susceptibility can be obtained by
comparing the Van Vleck term $\chi_{VV}\sim\frac{60}{7 \Delta E}=
5.7\times10^{-3}$ obtained from Equation~\ref{eq:2} to the
Curie-Weiss term $\chi_{CW}\sim
\mu_B^{-2}\mu_{eff}^{2}/(T-\Theta_p)$. For example with
$\mu_{eff}=0.845\mu_B$, at 300 K $\chi_{CW}= 2.6 \times10^{-3}$ and
at 100 K $\chi_{CW}= 9.9 \times10^{-3}$.

The effective magnetic moment derived from the Curie constant $C$
for the three films is $\mu_{\mathrm{eff}} = 0.45 \pm 0.1 \mu_B$ per
Sm ion, somewhat smaller than the free ion effective paramagnetic
moment of $g_J [J(J+1)]^{\frac{1}{2}} \mu_B = 0.845 \mu_B$, though
still larger than the value of $0.035\mu_B$ in the ferromagnetic
phase. The paramagnetic moment is understood by noting that the
$J=\frac{5}{2}$ level is decomposed by the cubic octahedral crystal
field into a doublet $\Gamma_7$ and a quartet $\Gamma_8$. Specific
heat data \cite{stutius1969tsh} quoted in \cite{moon1979mpo} report
that the $\Gamma_7$ sublevel is the ground state with a separation
of $225$~K to the $\Gamma_8$ sublevel for the bonding configuration
of SmN. We can therefore assume that at $30$~K only the ground
doublet is significantly populated and calculate an approximate
value of the low temperature susceptibility. The $\Gamma_7$ Kramers
doublet \cite{lea1962tro} can be described with the equivalent wave
functions $\vert\pm\frac{1}{2}\rangle$ for a fictitious spin
$S'=\frac{1}{2}$, for which we have calculated a Land\'e factor,
$g'$, of $\frac{10}{21}$. We obtain $\mu_{\mathrm{eff}}(\Gamma_7) =
[g'^2 \mu_B^2 S'(S'+1)]^{\frac{1}{2}} = 0.41 \mu_B$, in good
agreement with the experimental effective moment. It is important to
note that these results establish quite clearly that the entire Sm
population in the films participates in the paramagnetic Curie-Weiss
signal diverging at $T_C$, emphasising that the full population also
participates in the ferromagnetic order.

\section{Conclusion}

The present work gives strong evidence for a ferromagnetic state in
SmN. The magnetic moment in the ferromagnetic phase is an order of
magnitude smaller than in the paramagnetic state, confirming a
nearly-zero-moment ferromagnet below $27$~K. The magnetic behaviour
in the paramagnetic phase is in quantitative agreement with the
expected moments of Sm, showing the effects of both the excited
spin-orbit    state and the constraints imposed by the crystal
field. The reduced ferromagnetic moment is established quite clearly
by the experiments, but remains a theoretical challenge. The
near-zero moment ferromagnetic state in a semiconductor has clear
potential for various fundamental studies and devices involving
control of spin- and charge-degrees of freedom without the
perturbing effects of a fringe magnetic field. For device
applications it would clearly be interesting to investigate the
prospect of raising the Curie temperature of this material, either
by alloying or strain. However, there is also considerable
possibility that specific high technology spintronics devices that
run at $77$~K (and even He temperature) will be used in the future.

\begin{acknowledgments}
The MacDiarmid Institute is supported by the New Zealand Centre of
Research Excellence Fund and the research reported here by a grant
from the New Zealand New Economy Research Fund. C.M. is grateful to
the staff of the School of Chemical and Physical Sciences for their
hospitality, and thanks the financial support of the MacDiarmid
Institute and of the Royal Society of New Zealand.
\end{acknowledgments}

\end{document}